\documentstyle[sprocl,psfig]{article}

\def\simgt{\stackrel{>}{{}_\sim}}
\def\be{\begin{equation}}
\def\ee{\end{equation}}
\def\bear{\be\begin{array}}
\def\eear{\end{array}\ee}
\def\bea{\begin{eqnarray}}
\def\eea{\end{eqnarray}}

\newcommand{\msn}{m_{ \tilde{\nu} }  }
\newcommand{\msl}{m_{ \tilde{l}_L }  }
\newcommand{\msr}{m_{ \tilde{l}_R }  }

\begin{document}

\catcode`@=11
\newtoks\@stequation
\def\subequations{\refstepcounter{equation}%
\edef\@savedequation{\the\c@equation}%
  \@stequation=\expandafter{\theequation}
  \edef\@savedtheequation{\the\@stequation}
  \edef\oldtheequation{\theequation}%
  \setcounter{equation}{0}%
  \def\theequation{\oldtheequation\alph{equation}}}
\def\endsubequations{\setcounter{equation}{\@savedequation}%
  \@stequation=\expandafter{\@savedtheequation}%
  \edef\theequation{\the\@stequation}\global\@ignoretrue

\noindent}
\catcode`@=12
\phantom{.}
\vskip-1.5cm
\rightline{ IEM--FT--118/95}
\vskip0.8cm

\title{
NATURAL FLAVOUR MIXING IN THE MSSM AND  $\mu\rightarrow e,\gamma$}

\author{ B. DE CARLOS.}

\address{Department of Physics, Theoretical Physics, 1 Keble Road,\\
Oxford OX1 3NP, England}

\author{J.A. CASAS and J.M. MORENO}

\address{Instituto de Estructura de la Materia, \\
CSIC, Serrano 123, 28006 Madrid, Spain}

\maketitle\abstracts{
In the absence of any additional assumption it is natural to conjecture
that sizeable flavour-mixing mass entries, $\Delta m^2$, may appear in the
mass matrices of the scalars of the MSSM, i.e. $\Delta m^2\sim O(m^2)$.
This flavour violation can still be reconciled with the experiment
if the gaugino mass, $M_{1/2}$, is large enough to yield (through the
renormalization group running) a sufficiently small $\Delta m^2 / m^2$
at low energy. This leads
to a {\em gaugino dominance} framework (i.e. $M_{1/2}^2\gg m^2$),
which permits a remarkably model--independent analysis. We study this
possibility focussing our attention on the $\mu\rightarrow e,\gamma$ decay.
In this way we obtain very strong and general constraints, in particular
$\frac{M_{1/2}^2}{\Delta m}
\simgt 34\ {\rm TeV}$. }


\vspace{1cm}

It is well-known that FCNC processes are very sensitive tests to
physics beyond the standard model (SM) and, in particular, to
supersymmetric extensions of the SM (SSM) \cite{ellis82}.
Furthermore supersymmetry provides new direct sources of flavour violation,
namely the  possible (and even natural as we will see) presence of
{\em off-diagonal} terms (say generically $\Delta m^2$) in the squark
and slepton mass
matrices\cite{gabbi89,hagel94,choud95,brax94,barbi94,dimop95}.
In the present talk we will
focus all our attention on the constraints
on $\Delta m^2$ from the $\mu\rightarrow e,\gamma$ process because
they are very strong and, as we will see, their evaluation is remarkably
model--independent.

The minimal supersymmetric standard model (MSSM) is defined by
the superpotential, $W$ (from which the supersymmetric part of
the Lagrangian is readily obtained), and the
soft supersymmetry breaking terms coming
from the (unknown) supersymmetry breaking mechanism
\bea
-{\cal L}_{\rm soft}&=&\frac{1}{2}M_a\lambda_a\lambda_a +
\left(m_L^2\right)_{ij} \bar L_iL_j\ +\
\left(m_{e_R}^2\right)_{ij} \bar{e_R}_i {e_R}_j\
\nonumber \\
&+&\
\left(m_Q^2\right)_{ij} \bar Q_iQ_j\ +\
\left(m_{u_R}^2\right)_{ij} \bar {u_R}_i {u_R}_j\ +\
\left(m_{d_R}^2\right)_{ij} \bar {d_R}_i {d_R}_j
\label{nodiag}
\eea
$$
+\,  \left[A^u_{ij}h^u_{ij}Q_i H_2 {u_R}_j +
A^d_{ij}h^d_{ij}Q_i H_1 {d_R}_j
+ A^e_{ij}h^e_{ij}L_i H_1 {e_R}_j + B \mu H_1H_2 +{\rm h.c.} \right]\ ,
$$
where $i,j$ ($a$) are generation (gauge group) indices,
$\lambda_a$ are the gauginos,
and the remaining fields in the formula denote just their corresponding
scalar components in a standard notation. In the simplest version of the
MSSM the soft breaking parameters are taken as {\em universal} (at the
unification scale $M_X$). Then, the independent parameters of the theory are
\be
\label{inPar}
\mu,m,M_{1/2},A,B
\ee
(the rest of the parameters can be worked out demanding a correct
unification of the gauge coupling constants and correct masses for all
the observed particles). However this simplification is not at all a
general principle. In particular there is no theoretical argument
against non-vanishing off-diagonal
$(m^2_{ij})_{i\neq j}\equiv \Delta m^2_{ij}$ terms.
{}From the previous arguments,
it is natural to assume that these off-diagonal entries can
be {\em sizeable}, or even of the same order as the diagonal terms
\be
\label{sizedelt}
\Delta m^2 \sim O(m^2)\;\;.
\ee
Certainly, there are proposed mechanisms to avoid this, for example the
above-mentioned assumption of universality
\cite{nir93,dine93}. However one should wonder
whether, in the absence of any additional assumption, the perfectly possible
and even natural situation of eq.~(\ref{sizedelt}) could still be compatible
with the experimental data and, more precisely, with the present experimental
bound \cite{bolto88} on $\mu\rightarrow e,\gamma$
\be
\label{megexp}
BR(\mu\rightarrow e,\gamma)\le 5\times10^{-11}\;\;,
\ee
The expression of $BR(\mu\rightarrow e,\gamma)$ in the MSSM depends
on several low-energy quantities, namely $\tan\beta\equiv\langle
H_2\rangle/\langle H_1\rangle$, $\mu$, $A$, and the
spectrum of masses of sleptons and gauginos. These can be obtained
from the initial parameters
of the theory (see eq.~(\ref{inPar})) through the corresponding RGEs
(see ref. 11). On the other hand
the ratio $\Delta m^2/m^2$ will in
general be small at low energies (even if it is $O(1)$ at $M_X$),
provided that gaugino masses are bigger than scalar masses,
$M_{1/2}^2 \gg m^2$, because of the contribution
of the former in the RGEs of the diagonal parts of the latter, which is not
the case for the off-diagonal entries.
This is the reason why the RGEs have the potential to ``cure'' initial
sizeable values of $\Delta m$. Therefore, the assumption of naturally
large flavour mixing at $M_X$ leads us
necessarily to a $M_{1/2}^2 \gg m^2$ (``gaugino dominance'') \cite{dine90}
scenario,
where {\em all} the soft
breaking parameters are essentially determined {\em at low energies}
by the value of $M_{1/2}$ at $M_X$ independently of their initial
values. Note that all this does {\em not}
apply to the $\mu$ parameter, as it renormalises proportional to itself.
However, the further requirement of a correct
electroweak breaking fixes the value of $\mu$, giving us
the {\em whole} spectrum and other relevant low-energy quantities
(such as $A$ and $\tan\beta$) in terms of a {\em unique}
parameter $M_{1/2}$.
In particular, the values of $\tan \beta$ obtained in this framework
tend to be rather large (ranging from 11 to 26 as $M_{1/2}$ increases
from 150 GeV to 10 TeV. The gaugino dominance is a very interesting fact
that makes the subsequent analysis rather accurate and model--independent.

\vspace{0.3cm}
At lowest order, the $\mu \rightarrow e, \gamma$
process is induced by one--loop diagrams that
involve a flip of the leptonic flavour triggered by the slepton mixing,
besides the propagation of a neutralino or chargino (see ref. 11 for details).
Since the electron and muon Yukawa couplings are very suppressed,
only the gauge part of the couplings of the charginos and neutralinos
will play a role in the diagrams. In fact, one
important consequence of the gaugino dominance
framework is that for large enough values of $M_{1/2}$ the
neutralinos (charginos) are almost pure
neutral (charged) gaugino and higgsino.
Then, the relevant diagrams correspond to
bino ($\tilde{B}$) and wino ($\tilde{W}^{0}$, $\tilde{W}^{-}$) exchange.
We have evaluated {\em all} of them.
The expressions given in the previous literature
are either incomplete or not directly applicable to our case.

Although in principle all the diagrams can have a similar magnitude
(e.g. if we assume
$\Delta  m_{ \tilde{\nu}_e \tilde{\nu}_\mu }^2 \sim
 \Delta  m_{ \tilde{e}_L \tilde{\mu}_L }^2 \sim
 \Delta  m_{ \tilde{e}_R \tilde{\mu}_R }^2 $), in practice the bino diagram
is the dominant one. This comes from the coefficient of proportionality
$(A+\mu\tan\beta)$, that appears in its evaluation and turns out to be
very important in the gaugino dominance framework
due to the large $\tan\beta$ value.

The theoretical $BR(\mu \rightarrow e , \gamma)$ depends on two different
sets of parameters. First, the different masses involved in the game
($\msn^2, \; \msl^2, \; \msr^2,$ $\; M_{\tilde{B}}, \; M_{\tilde{W}}$)
and certain relevant  low-energy quantities $(A, \mu, \tan \beta)$.
Second, the three independent flavour-mixing mass entries:
$\Delta  m_{ \tilde{\nu}_e \tilde{\nu}_\mu }^2 , \;
 \Delta  m_{ \tilde{e}_L \tilde{\mu}_L }^2     , \;
 \Delta  m_{ \tilde{e}_R \tilde{\mu}_R }^2 $.
As explained in before, once we are working in the framework of gaugino
dominance, $M_{1/2}^2 \gg m^2 $, the first set is completely determined in
terms of the initial gaugino mass, $M_{1/2}$. Recall that we were led to
this framework by the mere assumption of naturally large flavour mixing at
$M_X$ (see eq.~(\ref{sizedelt})). The three flavour-mixing mass parameters,
however, remain independent.

The constraints on the MSSM from $BR(\mu \rightarrow e , \gamma)$
arise by evaluating the previous diagrams and comparing them with the
present experimental bound, eq.~(\ref{megexp}). We have illustrated this in
Fig.~1, where an overall mass-mixing parameter
$\Delta  m_{ \tilde{\nu}_e \tilde{\nu}_\mu }^2  \, = \,
 \Delta  m_{ \tilde{e}_L \tilde{\mu}_L }^2      \, = \,
 \Delta  m_{ \tilde{e}_R \tilde{\mu}_R }^2      \,\equiv \, \Delta  m^2 $
has been taken for simplicity. Then we have plotted
$BR(\mu\rightarrow e,\gamma)$ {\em vs}
$M_{1/2}$ for different values of $\Delta m$.
{}From this figure we can derive the maximum allowed value of $\Delta m$
(or, equivalently, the minimum allowed value of
$M_{1/2} / \Delta m$) for each value of
$M_{1/2}$. This is represented in Fig.~2 for four different cases:
{\em a)} $\Delta  m_{ \tilde{\nu}_e \tilde{\nu}_\mu }^2  \, = \,
 \Delta  m_{ \tilde{e}_L \tilde{\mu}_L }^2        \, = \,
 \Delta  m_{ \tilde{e}_R \tilde{\mu}_R }^2        \, \equiv \, \Delta  m^2 $;
{\em b)} only
$ \Delta  m_{ \tilde{e}_R \tilde{\mu}_R }^2       \, \ne \, 0$;
{\em c)} only
$ \Delta  m_{ \tilde{e}_L \tilde{\mu}_L }^2       \, \ne \, 0$
and
{\em d)} only
$ \Delta  m_{ \tilde{\nu}_e \tilde{\nu}_{\mu} }^2 \, \ne \, 0$,
which gives a complete picture of the results.
Notice that the {\em (d)} case is the less restrictive one.

The constraints are in general extremely strong. For
case {\em (a)}, which is the most representative one, the corresponding
curve can be approximately fitted by the simple
constraint
\be
\label{fit}
\frac{M_{1/2}^2}{\Delta m}\simgt 34\ {\rm TeV}
\ee
(similar equations can be written for the other curves). Under the
assumption of eq.~(\ref{sizedelt}), i.e. $\Delta m = O(m)$, the results
of Fig.~2 or eq.~(\ref{fit}) imply that, indeed, a very large
hierarchy between the scalar and gaugino masses is needed in order
to reconcile the theoretical and experimental results. This
gives full justification to our assumption of a gaugino dominance
framework once eq.~(\ref{sizedelt}) has been conjectured. For
example, for $M_{1/2}\sim 500$ GeV the assumption $\Delta m\sim m$
demands $M_{1/2}/\Delta m> 65$.
Actually, it is hard to think of a scenario where such a dramatical
hierarchy can naturally arise. Consequently, we can conclude at this
point that a naturally large flavour mixing, as that conjectured in
eq.~(\ref{sizedelt}), can hardly be reconciled with the experiment in
a natural way.

We can now summarize our work.
In the absence of any additional assumption it is natural to conjecture
that sizeable flavour-mixing mass entries, $\Delta m^2$, may appear in the
mass matrices of the scalars of the MSSM, i.e. $\Delta m^2\sim O(m^2)$.
This flavour violation can still be reconciled with the experiment
if the gaugino mass, $M_{1/2}$, is large enough to yield (through the
renormalization group running) a sufficiently small $\Delta m^2 / m^2$
at low energy. We have analyzed this possibility, focussing
our attention on the leptonic sector, particularly on the
$\mu\rightarrow e,\gamma$ decay, which is by far the FCNC process with
higher potential to restrict the value of the off-diagonal terms,
$\Delta m^2$. The results are the following:

\begin{enumerate}

\item The $\Delta m^2\sim O(m^2)$ conjecture automatically leads
to a {\em gaugino dominance} framework (i.e. $M_{1/2}^2\gg m^2$), where,
apart from $\Delta m^2$ itself, {\em all} the
relevant low-energy quantities (mass spectrum, $A$,
$\mu$, $\tan\beta$) are determined in terms of a unique parameter,
$M_{1/2}$.
This makes the subsequent analysis and results remarkably
model--independent.

\item The resulting constraints in the MSSM, obtained by comparing
the calculated $BR(\mu\rightarrow e,\gamma)$ with the experimental
bound, are very strong (see Figs.~1, 2 and  eq.~(\ref{fit}) ).
This makes, in our opinion, the natural flavour
mixing conjecture $\Delta m^2\sim O(m^2)$ extremely hard to be
reconciled with the experiment in a natural way.
Hence, $\Delta m/m$ should be
small already at the unification scale.

\end{enumerate}

\noindent

Finally, let us comment that the need of starting with small
$\Delta m/m$ can be satisfied in some theoretically well-founded
scenarios, which become favoured from this point of view.
 In particular,  we would like to stress that many
string constructions can be consistent with that requirement
\cite{kaplu93,brign94,decar93p}.
Other scenarios, however, can produce a larger non-universality
of the scalar masses, with potentially dangerous contributions to FCNC
processes \cite{brax94}. In any case, these non-universality effects
are to produce non-vanishing off-diagonal terms in the scalar
mass matrices once the usual rotation of fields to get diagonal
fermionic mass matrices is carried out. The
phenomenological viability of these physically relevant scenarios
undoubtedly deserves further investigation.


\newpage

\begin{figure}
\centerline{\vbox{
\psfig{figure=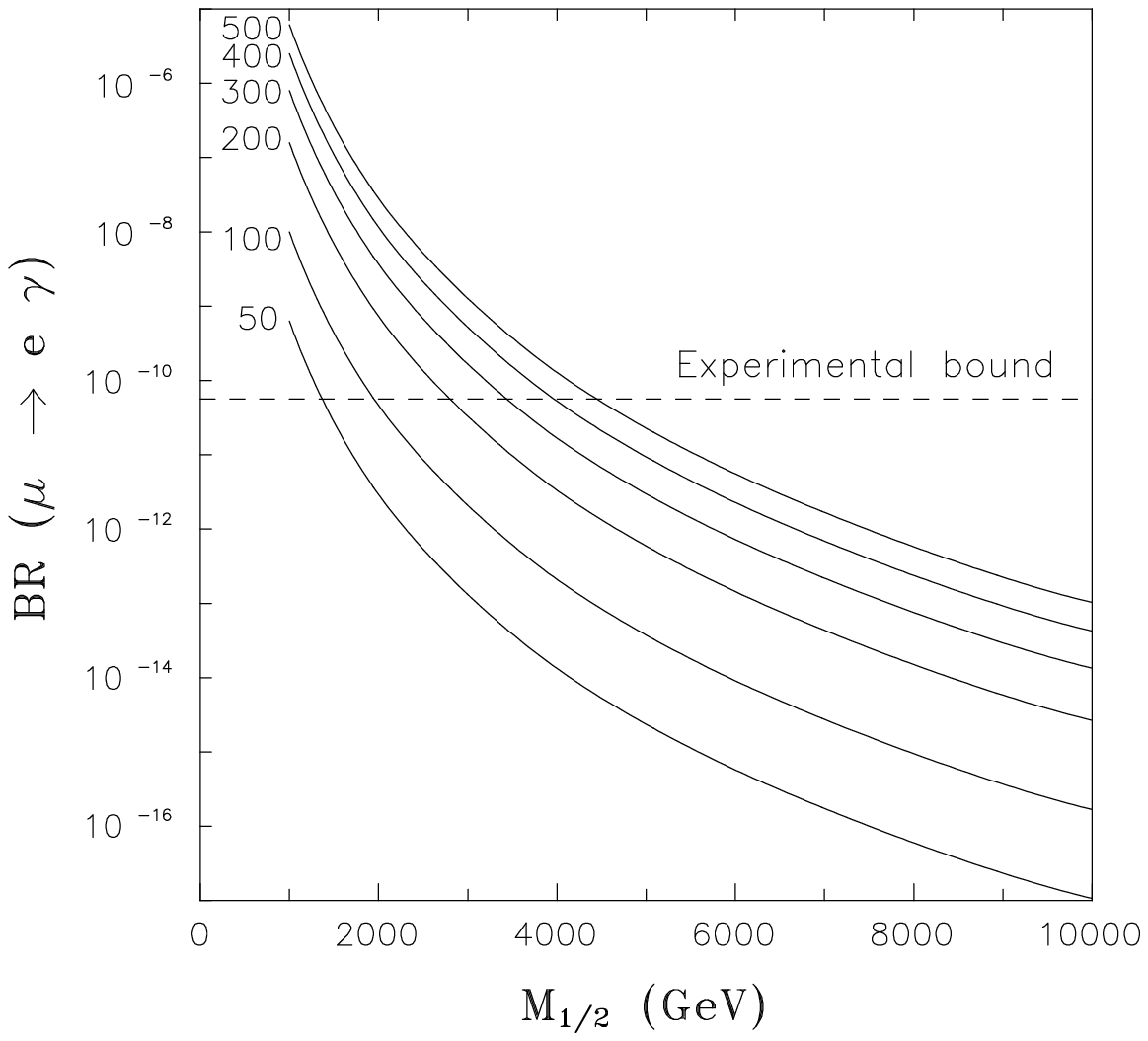,height=7.cm,bbllx=0.5cm,bblly=3.cm,bburx=13.5cm,bbury=14.5cm}
\caption{\tenrm\baselineskip=10pt Plot of $BR(\mu\rightarrow
e,\gamma)$ vs. $M_{1/2}$,
taking for simplicity $\Delta  m_{ \tilde{\nu}_e \tilde{\nu}_\mu }^2 =
\Delta  m_{ \tilde{e}_L \tilde{\mu}_L }^2  = \Delta  m_{ \tilde{e}_R
\tilde{\mu}_R }^2 \equiv  \Delta  m^2 $.
The different curves correspond to $\Delta m = $ 50, 100, 200, 300, 400,
500 GeV respectively.}
\psfig{figure=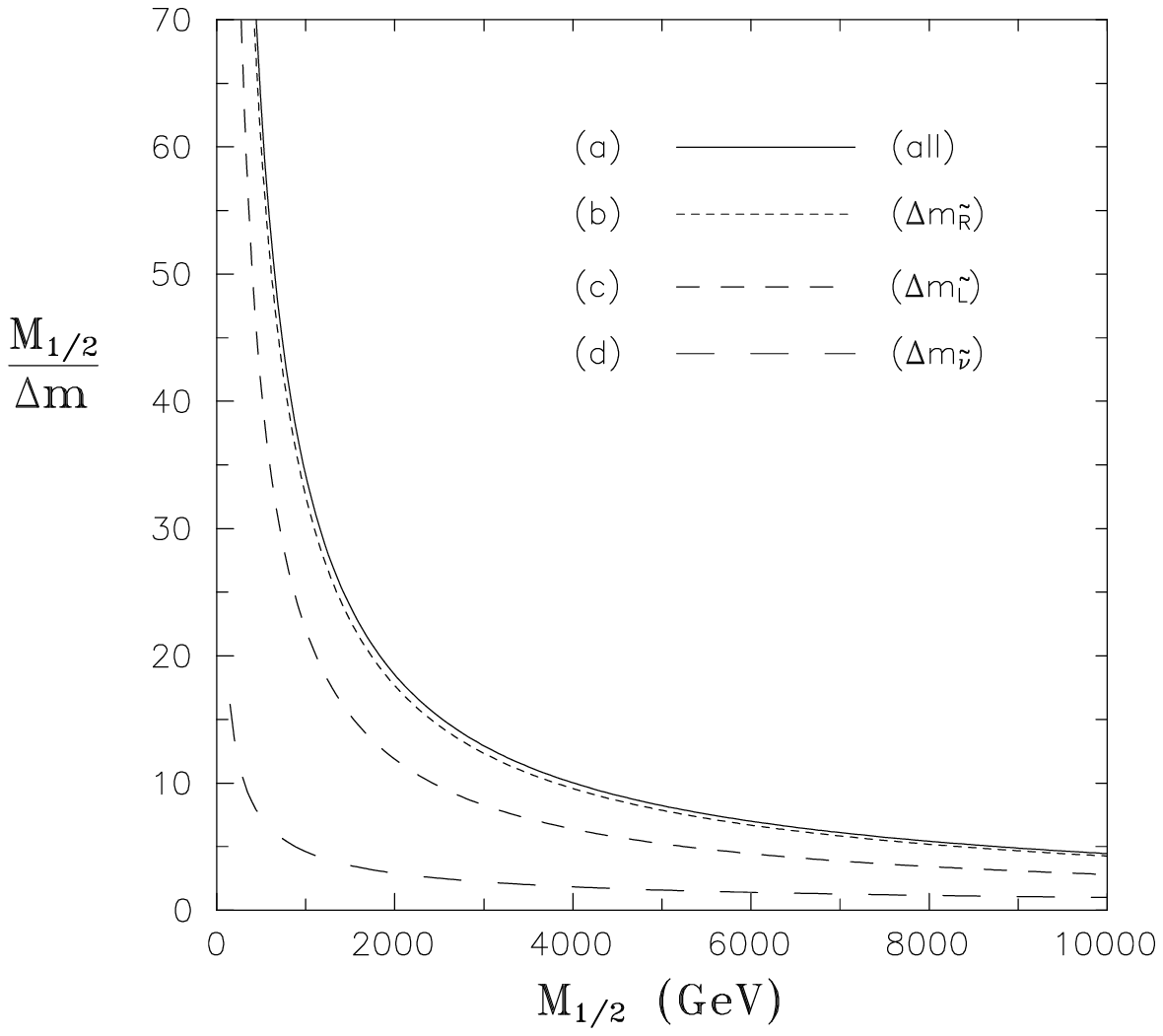,height=7.cm,bbllx=0.5cm,bblly=3.cm,bburx=13.5cm,bbury=14.5cm}
\caption{\tenrm\baselineskip=10pt
Plot of the {\em minimum} allowed
value of $M_{1/2}/\Delta m$ vs. $M_{1/2}$ in four different cases:
{\em a)} $\Delta  m_{ \tilde{\nu}_e \tilde{\nu}_\mu }^2  \, = \,
 \Delta  m_{ \tilde{e}_L \tilde{\mu}_L }^2        \, = \,
 \Delta  m_{ \tilde{e}_R \tilde{\mu}_R }^2        \, \equiv \, \Delta  m^2 $;
{\em b)} only
$ \Delta  m_{ \tilde{e}_R \tilde{\mu}_R }^2       \, \ne \, 0$;
{\em c)} only
$ \Delta  m_{ \tilde{e}_L \tilde{\mu}_L }^2       \, \ne \, 0$
and
{\em d)} only
$ \Delta  m_{ \tilde{\nu}_e \tilde{\nu}_\mu }^2 \, \ne \, 0$.}}}
\end{figure}

\end{document}